# Identifying the role that individual animals play in their social network[1]

David Lusseau* and M.E.J. Newman†

**ABSTRACT**

Techniques recently developed for the analysis of human social networks are applied to the social network of bottlenose dolphins living in Doubtful Sound, New Zealand. We identify communities and subcommunities within the dolphin population and present evidence that sex- and age-related homophily play a role in the formation of clusters of preferred companionship. We also identify brokers who act as links between subcommunities and who appear to be crucial to the social cohesion of the population as a whole. The network is found to be similar to human social networks in some respects but different in some others such as the level of assortative mixing by degree within the population. This difference elucidates some of the means by which the network formed and evolves.

---

[1] Submitted to *Ecology Letters*

The dolphin association data was collected by DL, Oliver J. Boisseau, Patti Haase, and Karsten Schneider for the long-term research program of the University of Otago- Marine Mammal Research Group which is directed by Steve Dawson and Liz Slooten. Data collection was funded by the New Zealand Whale and Dolphin Trust and Real Journeys Ltd. MEJN was funded in part by the US National Science Foundation under grant number DMS-0234188. Additional support to DL was provided by University of Aberdeen-Lighthouse Field Station.

*Lighthouse Field Station, University of Aberdeen, George Street, Cromarty, Ross-shire IV11 8YJ, Scotland. E-mail: d.lusseau@abdn.ac.uk

†Center for the Study of Complex Systems, University of Michigan, Ann Arbor, MI 48109. U.S.A. E-mail: mejn@umich.edu

## INTRODUCTION

Sophisticated tools for the study and analysis of social structure in human populations have been developed over the last half century (Wasserman & Faust 1994; Scott 2000). At the same time a variety of studies have revealed the structure of social networks in particular animal communities (Connor *et al.* 1999; McComb *et al.* 2001; Mitani *et al.* 2002; Lusseau 2003). Combining these resources, the analysis of animal social networks can offer substantial insights into the social dynamics of animal populations and possibly suggest new management strategies (Anthony & Blumstein 2000). Animal social networks are substantially harder to study than networks of human beings because animals do not give interviews or fill out questionnaires, and network data must be gathered by direct observation of interactions between individuals. Nonetheless, it has recently been possible to determine behaviourally meaningful measures of association in a number of species (Mitani *et al.* 2002; Lusseau 2003; Lusseau *et al.* 2003)). Early studies of animal social networks showed striking similarities to human networks (Connor *et al.* 1999; McComb *et al.* 2001; Mitani *et al.* 2002). We employ here a number of recently developed techniques to detect the role played by different individuals and categories of individuals in the cohesion of societies. We analyse the social network of a community of bottlenose dolphins living in Doubtful Sound, New Zealand and focus particularly on the clustering or community structure of the network and possible connections between this structure and attributes of the dolphins.

The network we study was constructed from observations of a community of 62 bottlenose dolphins (*Tursiops* spp.) over a period of seven years from 1994 to 2001 (Lusseau 2003). Nodes in the network represent the dolphins and ties between nodes represent associations between dolphin pairs occurring more often than expected by chance (Fig. 1). First, we dissect the network using a previously proposed clustering algorithm based on the calculation of betweenness scores, and extract the natural divisions in the dolphin community. Then we examine the relationship between these divisions and the sex and age of the dolphins. In the second part of the study we investigate the role played by different individuals in maintaining the cohesion of communities and of the whole network.

## COMMUNITY STRUCTURE AND ASSORTATIVE MIXING

Many methods for detecting communities within social networks have been described over the years (Scott 2000). Here we make use of one of the most recent, a method proposed by Girvan and Newman (Girvan & Newman 2002; Newman & Girvan 2004), which appears to be accurate and sensitive. The method finds natural divisions of networks into tightly knit groups by looking for the edges that run between groups. These edges are identified using a “betweenness'' measure which is a generalisation to edges of the vertex betweenness measure of Freeman (Freeman 1977). Edges with the highest scores by this measure are removed from the network, leaving behind the groups themselves. Two communities and four sub-communities were detected in the dolphin network (Fig. 1, Newman and Girvan 2004).

While many studies have been content with merely finding community structure within networks, we are here concerned not just with the fact that divisions exist within our network, but also with understanding how these divisions arise in this case. One mechanism for the formation of communities is homophily, or assortative mixing, the preferential association of individuals with others who are like them in some way (Newman 2002). In human societies individuals have been observed to associate along lines of race, gender, age, income, or nationality, to name but a few, giving rise to communities composed of individuals with similar profiles as measured by these factors.

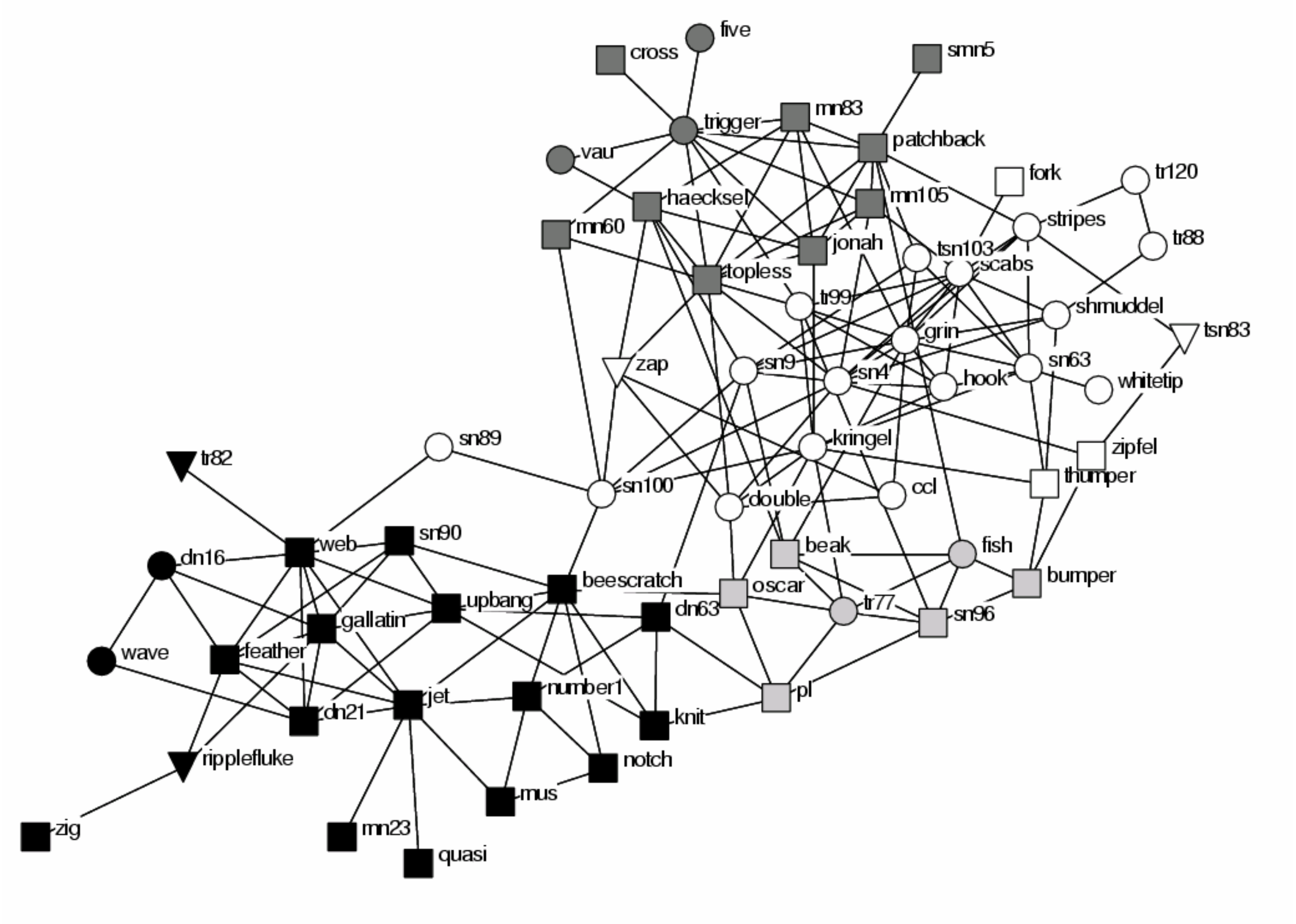

**Figure 1** Communities and sub-communities identified in the dolphin social network using the betweenness-based algorithm of Girvan and Newman (2002). Vertex colour indicates community membership: black and non-black vertices represent the principle division into two communities. Shades of grey represent sub-communities. Females are represented with circles, males with squares and individuals with unknown gender with triangles.

In our dolphin network, the sex of dolphins was known for all but three individuals from direct observations of genitalia using an underwater camera (Lusseau *et al.* 2003). This factor appears to play an important role in the definition of sub-communities (Fig. 1). This effect can be quantified using an “assortativity coefficient” (Newman 2003). Let $e_{ij}$ be the fraction of ties in the network that connect individuals of type *i* to individuals of type *j*. Then the assortativity coefficient is defined as

$$r = \frac{\sum_i e_{ii} - \sum_{ijk} e_{ik} e_{jk}}{1 - \sum_{ijk} e_{ik} e_{jk}}$$

This quantity equals 1 when we have perfect assortative mixing- all individuals associate solely with others of the same type as themselves- and zero when mixing is random. Partial mixing gives values between 0 and 1.

For example, in the present network there are 58 ties between males and males, 46 between females and females, and 44 between males and females, for a total of 159 ties altogether. This gives us $e_{mm}$ = 0.36, $e_{ff}$ = 0.29, and $e_{mf}$ = $e_{fm}$ = 0.28, where we have divided the male-female ties equally between $e_{mf}$ and $e_{fm}$ so that no tie appears in both

$$r = \frac{e_{mm} + e_{ff} - (e_{mm} + e_{mf})^2 - (e_{ff} + e_{fm})^2}{1 - (e_{mm} + e_{mf})^2 - (e_{ff} + e_{fm})^2}$$

This gives r = 0.346 ±0.053, where the error is calculated as described in Newman (2003). Hence, there is clear statistically significant assortative mixing by sex among the dolphin population, although the mixing is not as strong as some types of mixing seen in human societies (Newman 2003).

Assortative mixing by age is common in human social networks, and we can test for it also in dolphin networks, although our test is cruder than for mixing by sex because our data are poorer. We do not have exact figures for the ages of the dolphins in this study: bottlenose dolphins typically live several decades, and all the individuals included in the study were born before the start of the period of observation. We can make a crude estimate of age based on their size, their association with their mothers at the beginning of the study, and in the case of females whether they have been observed bearing calves during the study period (Lusseau *et al.* 2003). Also these dolphins, especially males, tend to accumulate scars during their lifetimes from fights with others and from shark attacks. The more scarred an individual is, the more likely it is that he or she is older. Based on a combination of all of these factors, individuals were divided into two groups, corresponding to older and younger dolphins, and assortativity was then measured according to this division. We find a value of r = 0.148 ±0.044 for mixing by age, considerably weaker than that for mixing by sex, although still statistically significant. Some mixing by age is common in many dolphin communities because female bottlenose dolphins tend to change their association patterns when they become pregnant, and spend more time with other pregnant females in so-called nursery schools (Wells *et al.* 1987). This segregation continues for some time after the calves are born, so that calves tend to be raised with others of their own age, thereby forming stronger bonds. This segregation of mothers was, however, not observed amongst the dolphins of Doubtful Sound (Lusseau *et al.* 2003) and thus we might expect to see a lower level of assortativity by age, as observed here. It would be interesting to compare our results for age assortativity with similar studies of other dolphin populations. Such a comparison might help us to quantify the role played by social bonding at an early age in shaping societies.

Assortative mixing by vertex degree, i.e., by the number of ties individuals have, is often observed in human social networks. Essentially all human networks are found to show positive assortative mixing by degree, the gregarious people tend to associate with other gregarious people and the hermits with other hermits (Newman 2003). Interestingly, the dolphin network studied here shows no such bias. Assortative mixing by degree can be quantified by calculating a simple Pearson correlation coefficient between the degrees of adjacent vertices in the network. For the present network this yields a value of $r = -0.044 \pm 0.080$, which is a null result. This appears to rule out some mechanisms of network evolution that are thought to be active in human societies. Both the triadic closure process, in which individuals tend to form ties to the friends of their friends (Banks & Carley 1996; Davidsen *et al.* 2002), and the preferential attachment process, in which individuals form ties to others with many ties (Barabasi & Albert 1999), are expected to produce assortative mixing by degree. These mechanisms also normally give rise to heavy-tailed degree sequences in networks (Barabasi *et al.* 1999; Dorogovtsev *et al.* 2000; Krapivsky *et al.* 2000; Davidsen *et al.* 2002), and the fact that no such a degree sequence is observed in the present network (Lusseau 2003) also argues for their absence here.

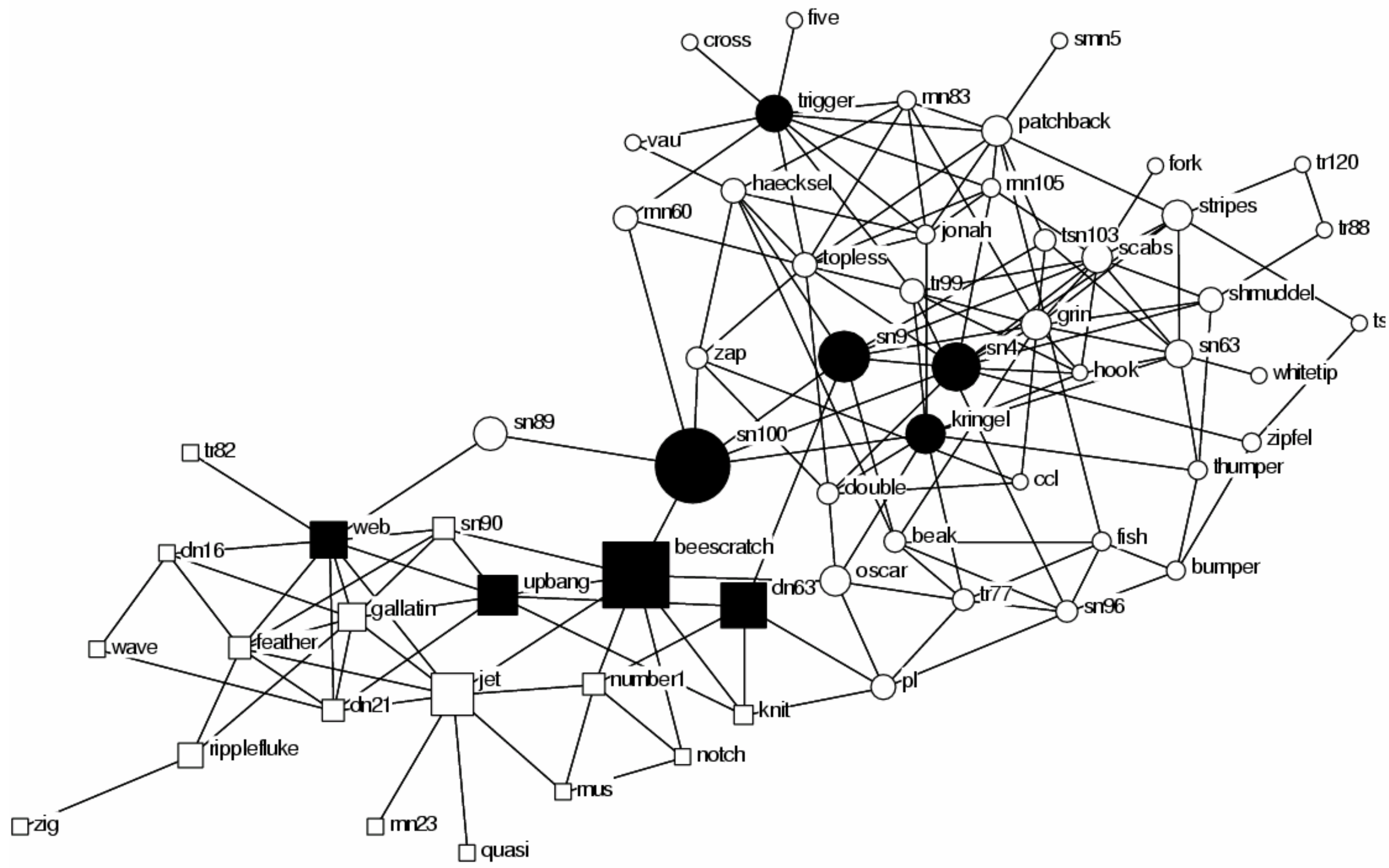

**Figure 2** The two communities present in the dolphin social network (squares and circles) with individuals with high betweenness values (greater than 7.32) represented by filled symbols. The size of the filled symbols is directly related to the betweenness of the vertex.

## CENTRALITY MEASURES AND THE ROLES OF INDIVIDUALS

Betweenness (Freeman 1979) is a measure of the influence of individuals in a network over the flow of information between others. The betweenness of a vertex $i$ is defined as the number of shortest paths between other pairs of vertices that pass through *i*. In the dolphin network, the vertices with highest betweenness fall, not surprisingly, on the boundary between the communities in the network (Fig. 2). The communities were defined by looking for edges with high betweenness, not vertices, but edge and vertex betweennesses tend to be correlated. The betweenness centrality thus tends to pick out boundary individuals who play the role of brokers between communities.

There is evidence, in fact, that a temporary disappearance of the dolphin denoted SN100, the individual with the highest betweenness centrality, led to the fission of the dolphin community of Doubtful Sound into two subgroups, the split occurring along the line of the community boundary identified in our community structure analysis. The recent reappearance of SN100 seems to be coinciding with a reunion of the two communities (Lusseau *et al.*, unpublished data). It appears therefore that betweenness is not merely a structural property in our network but is correlated with real social behaviours in the dolphin society; the highest betweenness individual was clearly playing an important role holding the community together.

Empirical evidence indicates that betweenness centrality may have a power-law distribution in many networks (Goh *et al.* 2002). We do not see such a distribution in the present network, but the betweenness distribution is strongly right-skewed, with a cumulative distribution that approximately follows an exponentially truncated power law (Fig. 3). The betweenness follows a power law approximately up to values around 7.32 (straight portion to left of figure) and then is sharply cut off above this value. Thus most individuals in the network have little influence over others in the sense associated with betweenness centrality, but a small proportion in the tail of the distribution are much more influential and may be regarded as key individuals who can control the flow of information in the society.

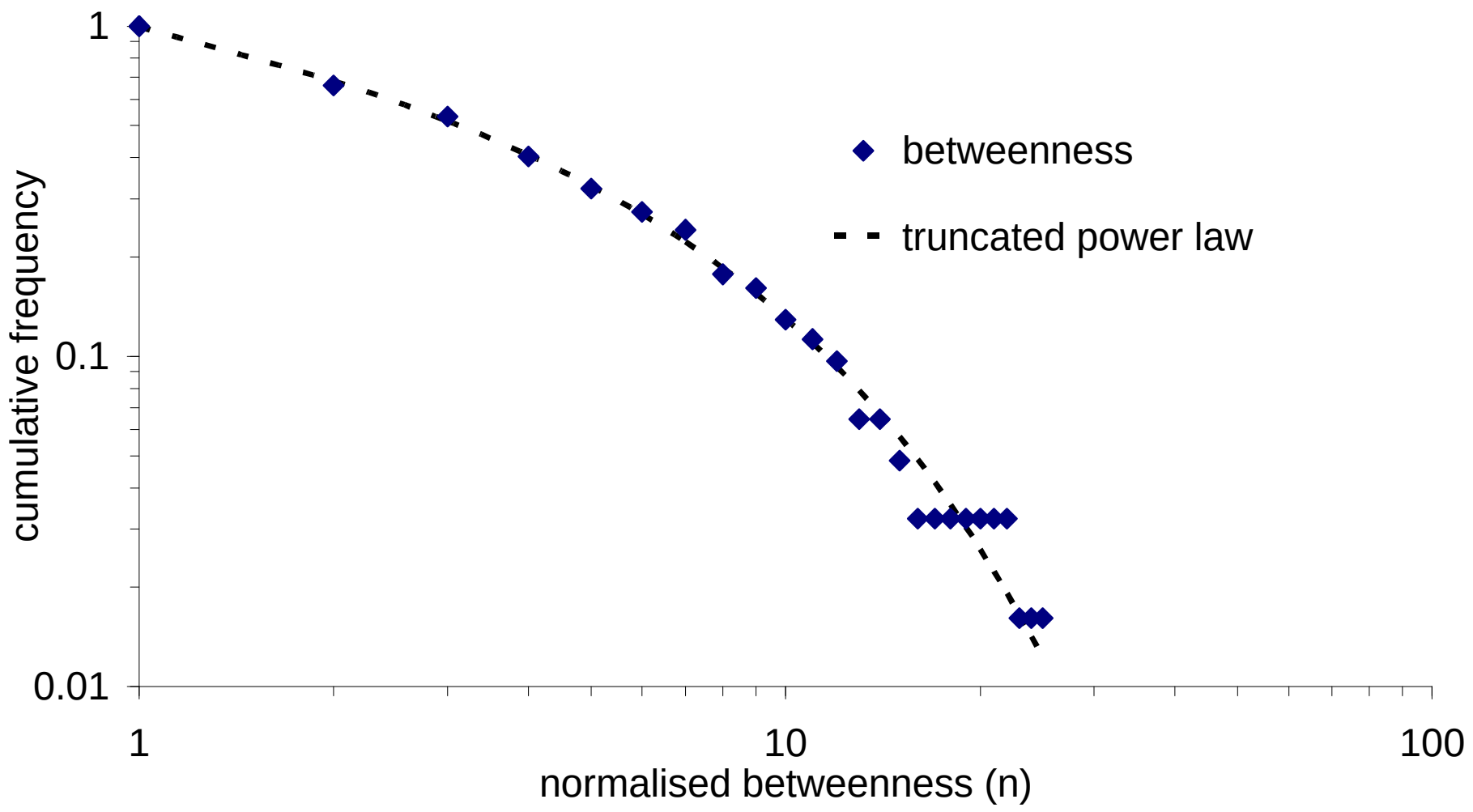

**Figure 3** Log-log plot of the cumulative frequency distribution of the betweenness scores of vertices. The line represents a truncated power-law fit to the data: $p(n) = an^{-b}e^{-\frac{n}{c}}$, where n is the betweenness value, p(n) is the cumulative frequency distribution of the betweenness, a (= 1.14) is a fitting constant, b (= 0.35) is the power law exponent, and c (= 7.32) is the point at which the curve departs from a power law distribution.

Curiously, despite the evident influence of SN100, the network is not highly susceptible to the removal of vertices with high betweenness. In many networks, the removal of high-betweenness individuals is a very effective way of destroying network connectivity (Holme *et al.* 2002). All vertices in our network are connected by some path which can be broken by removing vertices. However, unlike some other social networks, it does not disintegrate very fast when the high-betweenness vertices are removed. When removing vertices with highest betweenness one by one, the largest component shrinks slightly faster than it would with random removal, but not much faster (Fig. 4). This appears to indicate that the network has many redundant paths of communication between its individual members. A similar picture is also seen if we remove the vertices with highest degree (Fig. 4). The dolphin society appears to be quite robust in this respect to loss of its members as it had been shown previously (Lusseau 2003). The case of SN100 however suggests that this simple analysis may be misleading. It is certainly possible that some individuals are more important to the connectivity of the network than others and that their removal causes a disproportionate effect not immediately evident from a simple picture of the network.

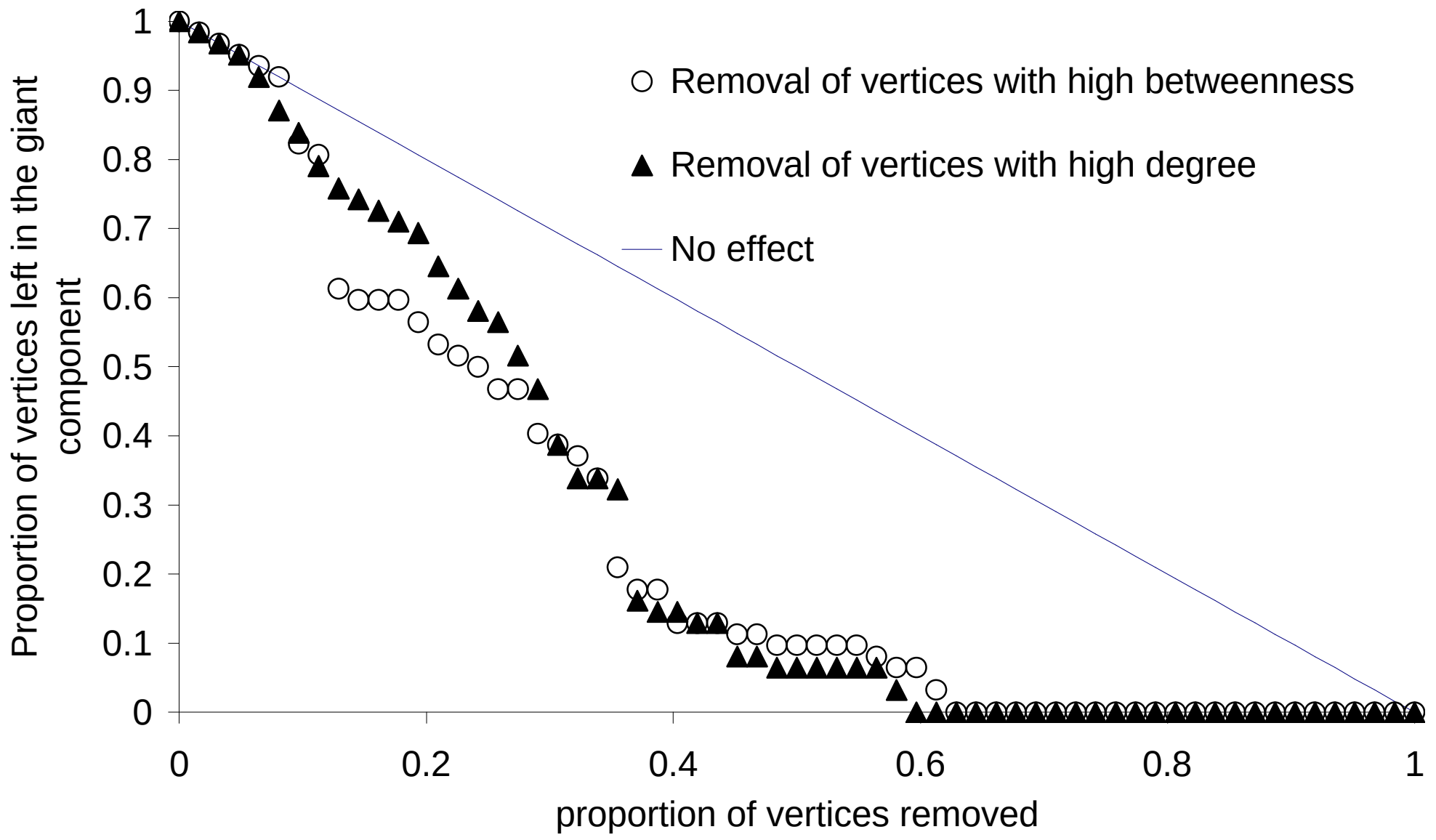

**Figure 4** Proportion of vertices left in the largest component of the network as vertices are removed. Circles represent results for vertices removed in order of decreasing betweenness centrality while triangles represent results for decreasing degree.

## CONCLUSIONS

In this paper we have described an analysis of the social network of a community of sixty-two bottlenose dolphins living in Doubtful Sound, New Zealand. The network was derived using observations of statistically significant frequent association between dolphin pairs. Using recently developed computer algorithms we have identified a number of sub-communities within the population, and we conjecture that these sub-communities may be a result of assortative mixing of dolphins by sex or age, although this conclusion must be considered tentative, since we have no good mechanistic understanding of how such mixing might arise in this case. The genetic relatedness of individuals is unknown and could also play an important role in community formation, as it does in other cetaceans, which tend to divide according to matrilineage (Connor *et al.* 1998). We also observe the existence of centralised "brokers" in the population, located at the boundaries between communities. Observations of the dynamics of the population as a whole suggest that these brokers may play a crucial role in maintaining the cohesiveness of the dolphin community. Overall, our results support the contention that association data can provide useful quantitative measures of social interaction among dolphins, and perhaps more generally in other animal communities.

More broadly, the techniques described here could improve our understanding of the effects of anthropogenic activities on animal populations. For instance, our findings suggest that the preservation of certain key individuals within a community may be crucial to maintaining its cohesion. Such information could help us to better target management actions by quantifying the important aspects of social structure in animal societies.